\documentclass[12pt]{article}
\usepackage{amsmath}
\usepackage{amssymb}
\usepackage{scicite}
\usepackage{graphicx}
\usepackage{times}
\usepackage{physics}
\usepackage{hyperref}
\newcommand{\llangle}{\left\langle}
\newcommand{\rrangle}{\right\rangle}

\newenvironment{sciabstract}{
\begin{quote} \bf}
{\end{quote}}

\title{Strange metallicity in the doped Hubbard model} 

\author
{Edwin W. Huang,$^{1,2\ast}$ Ryan Sheppard,$^{1}$ Brian Moritz,$^{2}$ Thomas P. Devereaux$^{2,3\ast}$\\
\\
\normalsize{$^{1}$Department of Physics, Stanford University, Stanford, CA 94305, USA}\\
\normalsize{$^{2}$Stanford Institute for Materials and Energy Sciences,}\\
\normalsize{SLAC National Accelerator Laboratory and Stanford University,}\\
\normalsize{Menlo Park, CA 94025, USA}\\
\normalsize{$^{3}$Department of Materials Science \& Engineering, Stanford University,}\\
\normalsize{Stanford, CA 94305, USA}\\
\\
\normalsize{$^\ast$To whom correspondence should be addressed;}\\
\normalsize{E-mail: edwinwhuang@gmail.com, tpd@stanford.edu}
}

\date{}

\begin{document}

\baselineskip24pt

\maketitle

\begin{sciabstract}
Strange or bad metallic transport, defined by incompatibility with conventional quasiparticle pictures, is a theme common to strongly correlated materials and ubiquitous in many high temperature superconductors. The Hubbard model represents a minimal starting point for modeling strongly correlated systems. Here we demonstrate strange metallic transport in the doped two-dimensional Hubbard model using determinantal quantum Monte Carlo calculations. Over a wide range of doping, we observe resistivities exceeding the Mott-Ioffe-Regel limit with linear temperature dependence. The temperatures of our calculations extend to as low as 1/40 of the non-interacting bandwidth, placing our findings in the degenerate regime relevant to experimental observations of strange metallicity. Our results provide a foundation for connecting theories of strange metals to models of strongly correlated materials.
\end{sciabstract}

Strongly correlated materials are renowned for their rich phase diagrams containing intertwined orders \cite{Fradkin2015,Keimer2015}. Difficulties associated with understanding emergence of these orders are largely rooted in the anomalous properties of the high temperature disordered phase. A most notable example is the DC resistivity in the normal state: numerous strongly correlated materials are strange or bad metals \cite{Emery1995}, where upon raising temperature sufficiently, resistivity exceeds the Mott-Ioffe-Regel (MIR) criterion with no sign of a crossover or saturation, signaling the absence of well-defined quasiparticles \cite{Gunnarsson2003,Hussey2004}. For many such systems, the resistivity is also characterized by linear temperature dependence up to the highest experimentally accessible temperatures. The incompatibility of these behaviors with conventional Fermi liquid theory poses a fundamental challenge to our understanding of strongly correlated electron systems. In particular, for the longstanding problem of high temperature superconductivity, it was recognized early on that transition temperatures in hole-doped cuprates are maximal where resistivity is most $T$-linear, suggesting that unconventional pairing is deeply connected to the nature of the strange metal.

The Hubbard model on a square lattice, containing only a local Coulomb interaction, is perhaps the most studied model of correlated electrons. While motivated in part by its believed relevance to cuprate superconductors, the model is generically important to the theoretical understanding of strong correlation effects due to its simple and plausibly realistic form. Lacking an analytical solution in two dimensions, the Hubbard model has been studied through a variety of numerical approaches primarily focusing on the nature of its ground state upon doping. Its transport properties remain relatively unexplored.

Here we demonstrate and study strange metallic transport in the normal state of the Hubbard model using determinantal quantum Monte Carlo (DQMC) calculations at finite temperatures \cite{BSS,White1989} combined with series expansions at infinite temperature \cite{Khait2016,Starykh1997,Lindner2010,Perepelitsky2016}. The Hubbard model Hamiltonian is $H = -\sum_{i j \sigma} t_{i j} c_{i \sigma}^\dagger c_{j \sigma} + U \sum_i c_{i \uparrow}^\dagger c_{i \uparrow} c_{i \downarrow}^\dagger c_{i \downarrow}$, where $c_{i \sigma}^\dagger$ is the creation operator for an electron on site $i$ with spin $\sigma$. The hopping energy $t_{i j}$ between sites $i$ and $j$ equals $t$ for nearest neighbors and $t'$ for next nearest neighbors. We choose $t'/t = -0.25$ and an intermediate interaction strength $U/t = 6$, and simulate $8 \times 8$ square clusters with periodic boundaries. Our simulations encompass a range of hole dopings from $p = 0$ to $p = 0.3$ and temperatures down to $T/t = 0.2$, or $1/40$ of the non-interacting bandwidth $W = 8 t$.

Our principal results are based on DQMC measurements of the current-current correlation function $\mathbf \Lambda(\tau) = \langle \mathbf j(\tau) \mathbf j \rangle$ where $\mathbf j = i \sum_{i j \sigma} t_{i j} (\mathbf r_i - \mathbf r_j) c_{i \sigma}^\dagger c_{j \sigma}$ is the current operator at momentum $\mathbf q = 0$ and $\tau$ is imaginary time. For the square clusters we study it is sufficient to consider only the $x x$ component of $\mathbf \Lambda(\tau)$. The optical conductivity $\sigma(\omega)$ relates to the imaginary time current-current correlation function through $\Lambda(\tau) = \int \frac{d\omega}{\pi} \frac{\omega e^{-\tau \omega}}{1-e^{-\beta \omega}} { \sigma}(\omega)$. We adopt the standard maximum entropy method of analytic continuation to extract the optical conductivity given DQMC measurements of the current correlator in imaginary time \cite{Jarrell1996,Gunnarsson2010}. Further details are provided in \cite{supp}, including data from larger cluster simulations indicating negligible finite size effects.

We first discuss the qualitative temperature dependence of optical conductivity (Fig.~1) for hole dopings $p=0$, $0.1$, and $0.2$. While we are concerned primarily with the metallic state of the doped system, it is important to establish the insulating nature of the undoped, half-filled model to verify strong correlation effects for our set of model parameters. The optical conductivity at half-filling, shown in Fig.~1A, demonstrates insulating behavior below roughly $T \sim t$, where cooling leads to a decreases of DC conductivity and formation of an optical gap. This behavior contrasts with the metallic properties of the doped case (Fig.~1B, C), where a Drude-like peak at zero frequency is present and the conductivity increases with lowering temperature. In the metallic regime, the increase in conductivity is primarily associated with narrowing of the $\omega = 0$ peak. Below $T \sim t$, relatively little spectral weight is transferred to or from the Hubbard peak at $\omega \approx U = 6 t$, which contains roughly the same spectral weight over a decade of temperature.

The metallic behavior at high temperatures is markedly distinct. For $T \gtrsim 1$, the optical conductivity and its temperature evolution are similar for all dopings, including half-filling. Broad peaks are present at $\omega = 0$ and $\omega \approx U = 6 t$. In this high temperature regime, the spectral weight in both peaks scale together when varying temperature. In contrast to the lower temperature metallic regime, here the width of the $\omega = 0$ peak does not evolve with temperature and the overall profile of the optical conductivity remains fixed.

Having explored the qualitative doping and temperature trends of the optical conductivity, we now focus on the Hubbard model's DC transport properties. The resistivity in natural units of $\hbar/e^2$ is plotted versus temperature in Fig.~2. The Mott-Ioffe-Regel (MIR) limit tends to be of order unity in natural units. Evidently in our data, no saturation related to the MIR criterion is present. In particular the resistivity for lightly doped systems significantly exceeds the MIR limit even at our lowest accessible temperature.

A clear distinction is present between temperatures below and above $T \sim 1.5$. As discussed previously, in the half-filled model, this temperature scale marks an onset of insulating behavior. In Fig.~1, we additionally saw that in the doped, metallic cases, $T \sim 1.5$ separates two regimes of qualitatively different temperature dependences in the optical conductivity. Here in Fig.~2, we see that the high and low temperature regimes differ also in the temperature and doping dependence of DC resistivity. While both regimes display $T$-linear resistivity, only at low temperatures $T \lesssim 1$ is there significant doping dependence to the resistivity. Going from $p=0.1$ to $p=0.3$, the temperature coefficient of resistivity decreases by roughly a factor of 3 for low temperatures while remaining nearly constant for $T \gtrsim 2$. For all considered dopings, the resistivity appears $T$-linear and uninfluenced by MIR, thus indicating that strange metallic transport is present through a significant portion of the Hubbard model's phase diagram.

To delineate the relevance of model calculations to material physics, it is instructive to consider the infinite temperature limit. For a generic nonintegrable model with a bounded energy spectrum, it is expected that $T \sigma(\omega)$ converges to a limit for temperatures above the largest energy scales of the model \cite{Mukerjee2006}, namely the ultra-high temperature limit. An immediate consequence is that large, linear-$T$ resistivity violating the MIR limit is ensured for sufficiently high temperature. While such behavior nominally reflects bad metallic transport, it is less relevant to experimental realizations of bad metals: generally both bad metals and saturating metals showcase their behaviors at temperatures significantly smaller than the Fermi temperature or interaction energy scales. In our calculations of the Hubbard model, we have seen that properties expected in the ultra-high temperature limit extend down to $T \sim 2$ before crossing over to a low temperature regime with distinct properties. The fact that the Hubbard model already violates MIR and displays $T$-linear resistivity in this low temperature regime suggests that its bad metallic transport is of a similar nature to that in strongly correlated materials.

Besides analyzing analytically continued optical conductivity, DC transport properties may be estimated through imaginary time proxies: simple functions of the imaginary time current correlator that converge to the true DC resistivity in low temperature limit. Intuitively, one expects low frequency properties to be most strongly related to data at large imaginary times. Specifically, $\tau = \beta/2$ is the ``largest'' imaginary time (since $\Lambda(\beta-\tau) = \Lambda(\tau)$). We first consider the proxy $\rho_1 = \pi T^2 \Lambda(\beta/2)^{-1}$, where $\Lambda(\beta/2) = \int d\omega f(\omega) \sigma(\omega)$. $f(\omega) = \frac{\omega}{2 \pi} / \sinh(\beta \omega / 2)$ is a bell-shaped function with width approximately $8 T$ that becomes a delta function for $T \to 0$ \cite{Trivedi1996}. $\rho_1$ thus approaches the true DC resistivity if the optical conductivity is featureless over the width of $f(\omega)$. In Fig.~1, we have seen that the zero frequency peak can be sharper than $8 T$, especially with increased doping. Due to this, $\rho_1$, plotted in Fig.~3A, deviates from the analytically continued data of Fig.~2.

The shortcomings of $\rho_1$ can be compensated by incorporating information of the curvature of the current correlator at $\tau=\beta/2$ \cite{Lederer2017}. In particular, $\rho_2 = \Lambda''(\beta/2)/(2 \pi \Lambda(\beta/2)^2)$ provides a more robust estimate of resistivity when the Drude-like peak is more narrow than $8 T$. As an example, if the optical conductivity consists of a Lorentzian peak at $\omega=0$ with width $\Gamma$, the ratio of the proxy to the DC resistivity ranges from $\rho_2/\rho_{DC} = 1$ for $\Gamma \gg T$ to $\rho_2/\rho_{DC} = 1/2$ for $\Gamma \ll T$. Plotting $\rho_2$ for our DQMC data in Fig.~3B, we see that $\rho_2$ captures many of the same features present seen in Fig.~2. While there may be differences in the precise value, in part due to limitations of this simple proxy, the trends and the decrease of the temperature coefficient with doping compare well with analytically continued results and corroborates the presence of strange metallicity in the Hubbard model.

To further analyze transport properties of the Hubbard model, we consider the Nernst-Einstein relation, which connects conductivity to charge compressibility and diffusivity: $\sigma = \chi \mathcal{D}$. In the context of correlated materials, since compressibility is nearly constant at experimentally relevant temperatures, the $T$-linearity of resistivity derives from the diffusivity, which has been argued to be a more fundamental transport property \cite{Hartnoll2015,Hartman2017}. In Fig.~4A, we plot the inverse compressibility, obtained in DQMC without analytic continuation. Qualitatively similar trends in doping dependence are present in the resistivity and inverse compressibility, which are somewhat cancelled out when combined to form the diffusivity (Fig.~4B). At high temperatures, since both resistivity and inverse compressibility scale linearly in temperature, the inverse diffusivity approaches a constant. Conversely at low temperatures, the compressibility approaches a limiting constant value. We thus see in Fig.~4 that the temperature dependence of resistivity crosses over from being dominated by compressibility \cite{Kokalj2017} to being controlled by diffusivity when lowering temperature. Interestingly, similar crossover behavior has been observed in a recent study of an extended Hubbard model in $t/U \to 0$ limit \cite{Mousatov2018}.

The presence of strange metallicity in the Hubbard model at temperatures small compared to the energy scales of model parameters provides promising evidence that the fundamental physics of correlated materials may be approached through studying simplified model Hamiltonians. In this regard we view thorough numerical results as presented here to be an important benchmark for testing theoretical descriptions of strange metals and approximate approaches to the Hubbard model \cite{Prushke1995,Bergeron2011,Deng2013,Xu2013}. A recent development involves measurement of transport properties in the Hubbard model via cold atoms experiments \cite{Xu2016,Brown2018,Nichols2018}, with broadly similar findings to our results. Both in this field and in finite temperature numerical approaches, studying the normal state down to temperatures proximate to ordering temperatures for superconductivity and other emergent phases remains a major challenge.

While ground state calculations of the Hubbard model have revealed intertwined orders with remarkable analogies to experimental phase diagrams \cite{Jiang2018,Zheng2017,Huang2018}, important questions remain concerning their emergence from the normal state. Controlled approaches to the Hubbard model at finite temperature, such as our DQMC calculations where there is a sign problem, currently are unable to directly access these phases.
Whether superconductivity in the Hubbard model follows directly from the strange metal as temperatures are lowered, or if coherent quasiparticles may emerge in between the strange metal and the ground state, remain intriguing open questions. Answers may be found through extending our measurements of dynamical quantities including resistivity, by developing new numerical techniques or via improved quantum simulations.


\bibliographystyle{Science}

\section*{Acknowledgments}
We acknowledge helpful discussions with Erez Berg, Luca Delacr\'etaz, Sean Hartnoll, Steve Kivelson, Yoni Schattner, and Jan Zaanen.
This work was supported by the U.S.~Department of Energy (DOE), Office of Basic Energy Sciences, Division of Materials Sciences and Engineering. Computational work was performed on the Sherlock cluster at Stanford University and on resources of the National Energy Research Scientific Computing Center, supported by the U.S.~DOE under Contract No.~DE-AC02-05CH11231.
Data supporting this manuscript are stored on the Sherlock cluster at Stanford University and are available from the corresponding author upon request. Source code for the simulations, including the MaxEnt analytic continuation code, is available at \url{https://github.com/edwnh/dqmc}.

\section*{Supplementary materials}
Materials and Methods\\
Supplementary Text\\
Figs.~S1 to S11\\
Table S1\\
References \textit{(31-33)}

\clearpage

\begin{figure}
\centering
\includegraphics{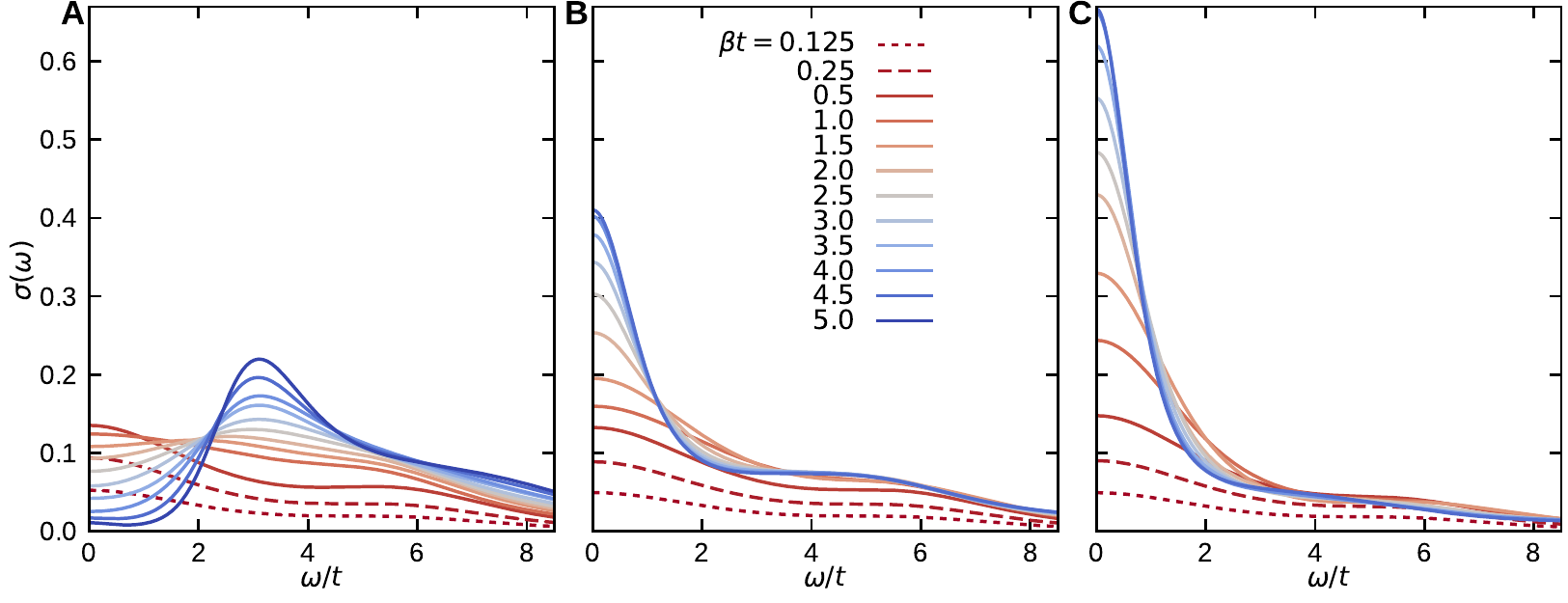}
\caption{{\bf Optical conductivity of the Hubbard model.} Optical conductivity obtained through DQMC and MaxEnt analytic continuation for the Hubbard model with parameters $U/t=6$, $t'/t=-0.25$. Hole doping level is $p=0.0$ ({\bf A}), $0.1$ ({\bf B}), and $0.2$ ({\bf C}). Simulation cluster size is $8 \times 8$; see \cite{supp} for comparison against simulations on larger clusters.}
\label{fig:f1}
\end{figure}

\begin{figure}
\centering
\includegraphics{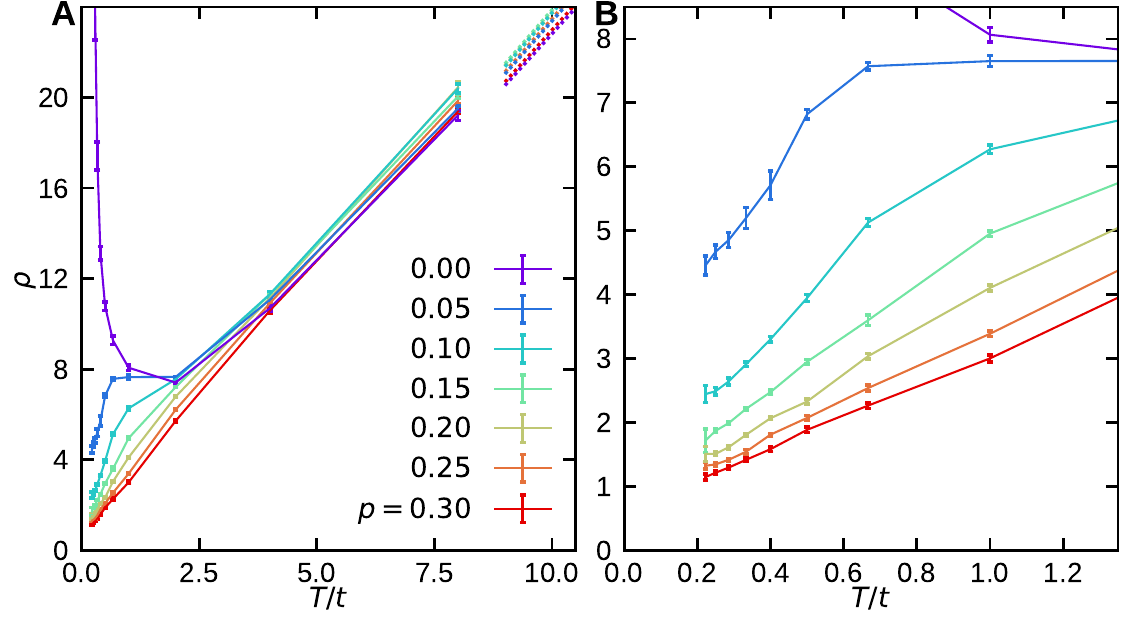}
\caption{{\bf DC resistivity extracted by analytic continuation.} ({\bf A}) DC resistivity as a function of temperature and hole doping, obtained from analytically continued optical conductivity as shown in Fig.~\ref{fig:f1}. Solid lines through DQMC data points are guides to the eye. Dotted lines are results from moments expansions up to 18th order in the high temperature limit \cite{supp}. ({\bf B}) Close-up view of the lowest temperature data of (A). Errorbars represent random sampling errors, determined by bootstrap resampling \cite{supp}.}
\label{fig:f2}
\end{figure}

\begin{figure}
\centering
\includegraphics{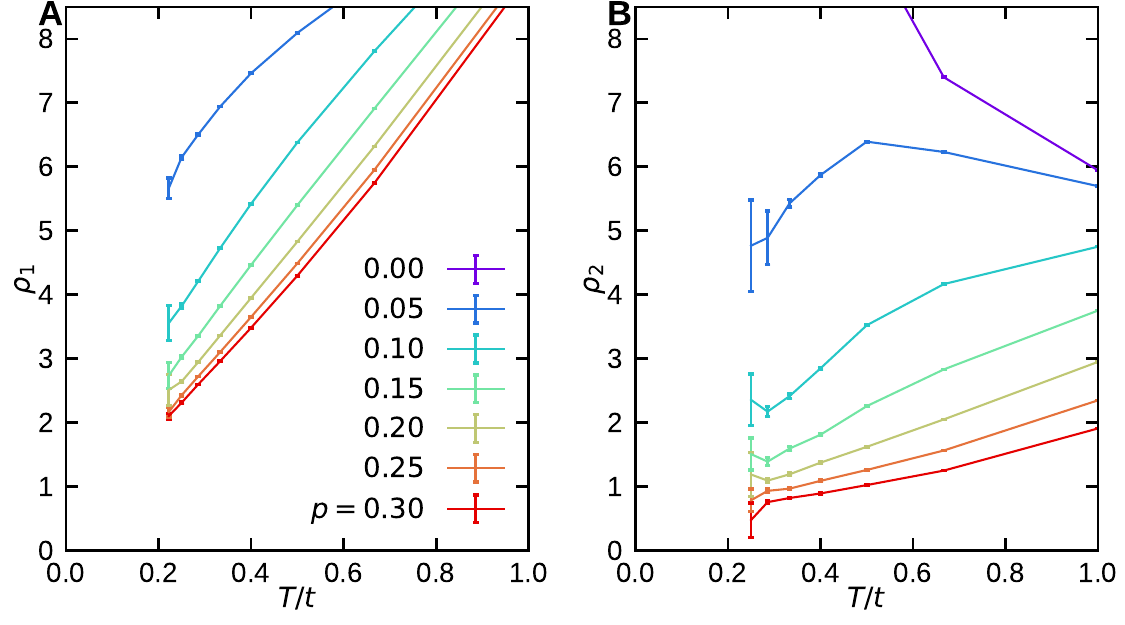}
\caption{{\bf DC resistivity via imaginary time proxies.} Proxies of DC resistivity $\rho_1 = \pi T^2 / \Lambda(\beta/2)$ ({\bf A}) and $\rho_2 = \Lambda''(\beta/2)/(2 \pi \Lambda(\beta/2)^2)$ ({\bf B}). Gray crosses correspond to data from a $12 \times 12$ simulation at $p=0.2$ hole doping. Errorbars are $\pm$ one standard error of mean, determined by bootstrap resampling.}
\label{fig:f3}
\end{figure}

\begin{figure}
\centering
\includegraphics{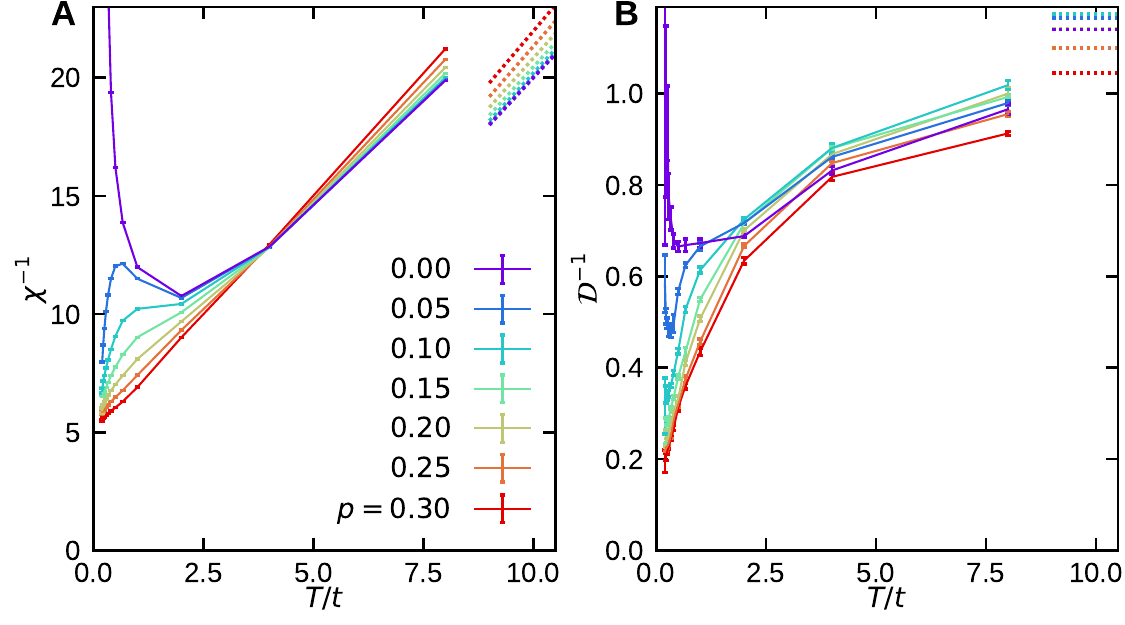}
\caption{{\bf Compressibility and diffusivity.} ({\bf A}) Inverse charge compressibility $\chi^{-1} = \left(\frac{\partial \langle n \rangle}{\partial \mu}\right)^{-1}$ calculated by DQMC simulations (without analytic continuation). Solid lines are guides to the eye; dotted lines are the high temperature limit $\chi = \frac{1-p^2}{2T}$. ({\bf B}) Inverse diffusivity obtained by applying the Nernst-Einstein relation $\sigma = \chi \mathcal{D}$ to the data of (A) and Fig.~\ref{fig:f2}.}
\label{fig:f4}
\end{figure}

\clearpage
\renewcommand{\thefigure}{S\arabic{figure}}
\setcounter{figure}{0}

\section*{Supplementary Materials}

\subsection*{Methods}

\noindent\underline{Hubbard model}

The Hubbard model Hamiltonian is
\begin{equation}
H = - \sum_{i j \sigma} t_{ij}\, c_{i\sigma}^\dagger c_{j\sigma} + U \sum_i n_{i\uparrow} n_{i\downarrow} - \mu \sum_{i\sigma} n_{i\sigma}
\end{equation}
where $c_{i\sigma}^{\dagger}$ ($c_{i\sigma}$) creates (annihilates) an electron with spin $\sigma$ at site $i$; $n_{i\sigma} = c_{i\sigma}^{{\dagger}} c_{i\sigma}$, the hopping $t_{ij}$ is equal to $t$ for nearest neighbors and $t'$ for next nearest neighbors, $U$ is the on-site repulsive Coulomb interaction, and the chemical potential $\mu$ controls the doping level.

\noindent\underline{Determinantal quantum Monte Carlo (DQMC)}

We perform DQMC simulations on the Hubbard model \cite{BSS,White1989} with parameters $U = 6$ and $t' = -0.25$. The chemical potential is tuned to achieve the desired doping level to within an accuracy of $O(10^{-4})$. The imaginary time interval $[0,\beta]$ is discretized into steps of at most $0.1$, resulting in negligible Trotter errors for our simulations. We consider cluster sizes of $8 \times 8$ in the main text and also consider larger clusters $12 \times 8$, $12 \times 12$, and $16 \times 8$ to investigate finite size effects.

To ensure numerical stability in computing the equal-time Green's functions, we use the prepivoting stratification algorithm as described in \cite{Tomas2012}, allowing up to 10 matrix multiplications before performing a QR decomposition. The unequal time Green's functions are constructed using the Fast Selected Inversion algorithm described in \cite{Jiang2016}, with blocks corresponding to the product of matrices from 10 time steps.

We generally run $500$ independently seeded Markov chains with $2 \times 10^6$ spacetime sweeps each, giving a total of $10^9$ sweeps for each doping and temperature. Fewer total sweeps ($100 \times 10^5$) are used in the high temperature simulations, where statistics tend to be better behaved. In all cases unequal time measurements are performed on every other sweep.

\noindent\underline{Analytic continuation of imaginary time data}

We perform maximum entropy analytic continuation (MaxEnt) \cite{Jarrell1996} to extract the optical conductivity from imaginary time current correlation data measured in DQMC. We use the classic formulation of MaxEnt with Bryan's algorithm for optimization \cite{Jarrell1996}. For the choice of model function, we use an annealing procedure where spectra from higher temperatures is used as model functions for lower temperature data. For the highest temperature data $T/t = 8$ or $\beta t = 0.125$, model functions come from the infinite temperature moments expansion discussed below in the Supplementary text.

\noindent\underline{Error analysis}

Sampling errors in our data may be estimated via bootstrap resampling of the bins of Moute Carlo data (each bin corresponds to an independently seeded Markov chain). Systematic errors stem from finite size effects and from the limitations of analytic continuation. All of these issues are discussed in greater detail below. Our analysis shows that our data have sufficiently small uncertainties to support our conclusions.

\subsection*{Supplementary text}

\noindent\underline{Cluster size dependence}

Generally, finite size effects can be expected to be minor at the temperatures of our calculations, since correlation lengths are small (e.g. 1-2 unit cells for spin-spin correlations). We explicitly demonstrate the absence of significant finite-size effects by considering simulations with size $12 \times 8$, $12 \times 12$, and $16 \times 8$. Due to the considerable computational expense of larger cluster simulations, especially in the presence of a sign problem, we focus on the intermediate hole doping $p=0.2$.

\begin{figure}
\centering
\includegraphics{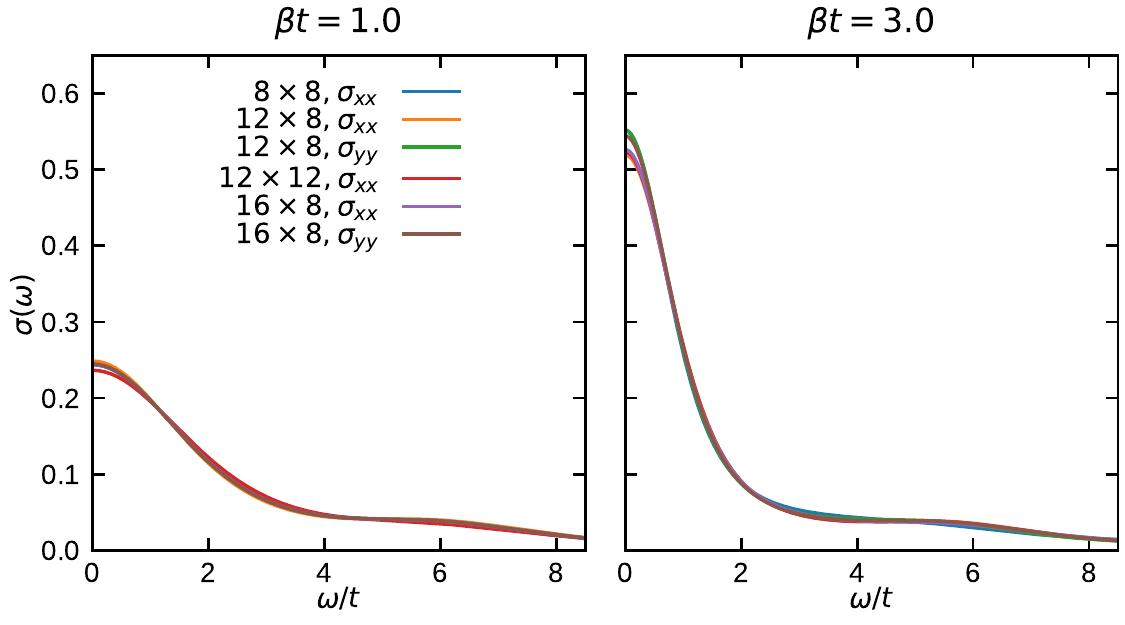}
\caption{Optical conductivity for $U/t=6$, $t'/t=-0.25$, $p=0.2$ and different cluster sizes.}
\label{fig:sigma_size}
\end{figure}

\begin{figure}
\centering
\includegraphics{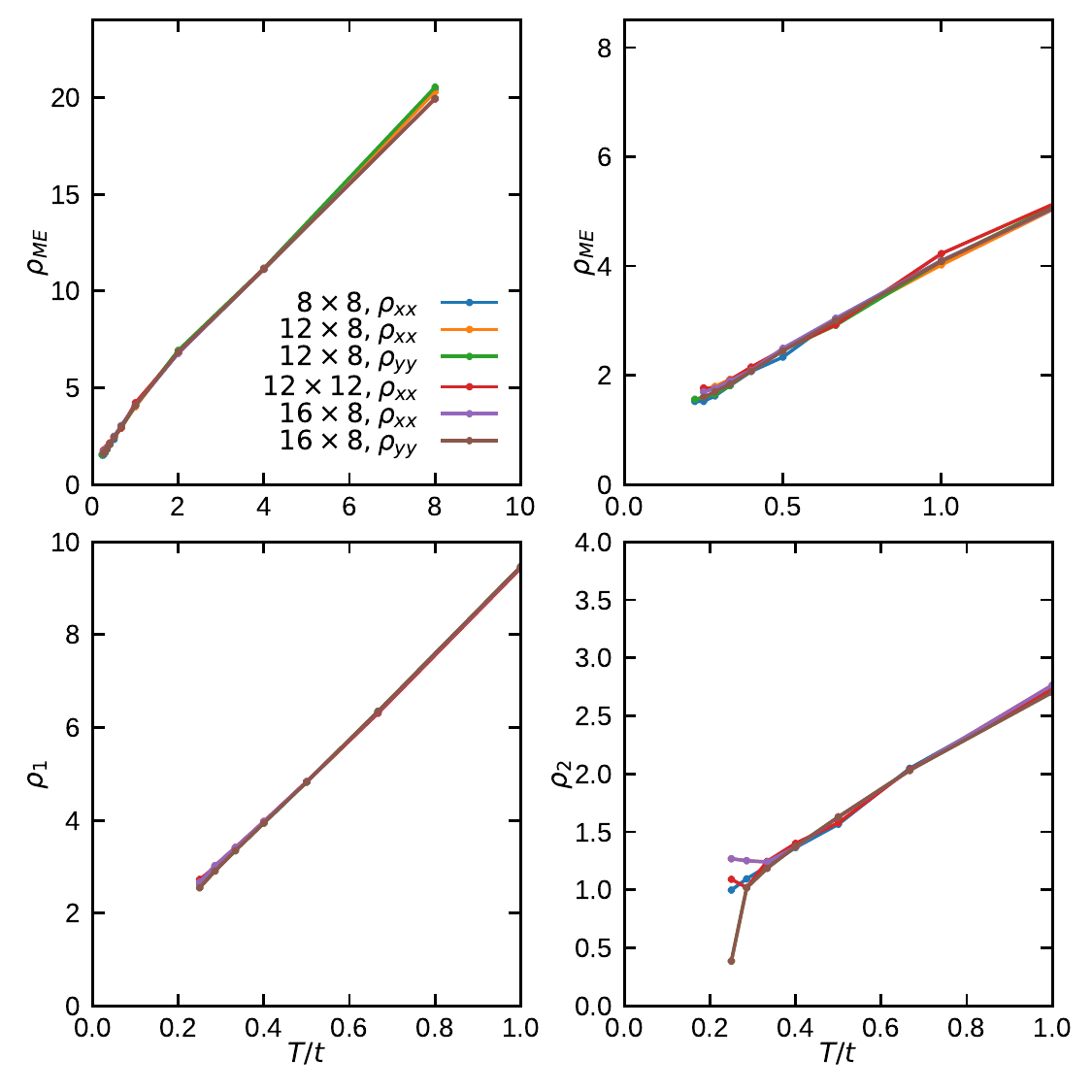}
\caption{DC resistivity for $U/t=6$, $t'/t=-0.25$, $p=0.2$ and different cluster sizes, evaluated through MaxEnt (top) and through the proxies (bottom) described in the main text}
\label{fig:rho_size}
\end{figure}

In Fig.~\ref{fig:sigma_size} we plot the optical conductivity for these three larger cluster sizes together with the $8 \times 8$ data displayed in Fig.~1C of the main text. For the rectangular clusters, we show both $\sigma_{x x}$ and $\sigma_{y y}$. In Fig.~\ref{fig:rho_size} we also show the temperature dependence of resistivity as obtained through MaxEnt and through the proxies $\rho_1$ and $\rho_2$ as discussed in the main text. Evidently the data in all cases are quantitatively similar for the larger clusters, thus indicating that the $8\times8$ cluster data presented in the main text are void of significant finite size effects.

\noindent\underline{Hubbard model parameter dependence}

\begin{figure}
\centering
\includegraphics{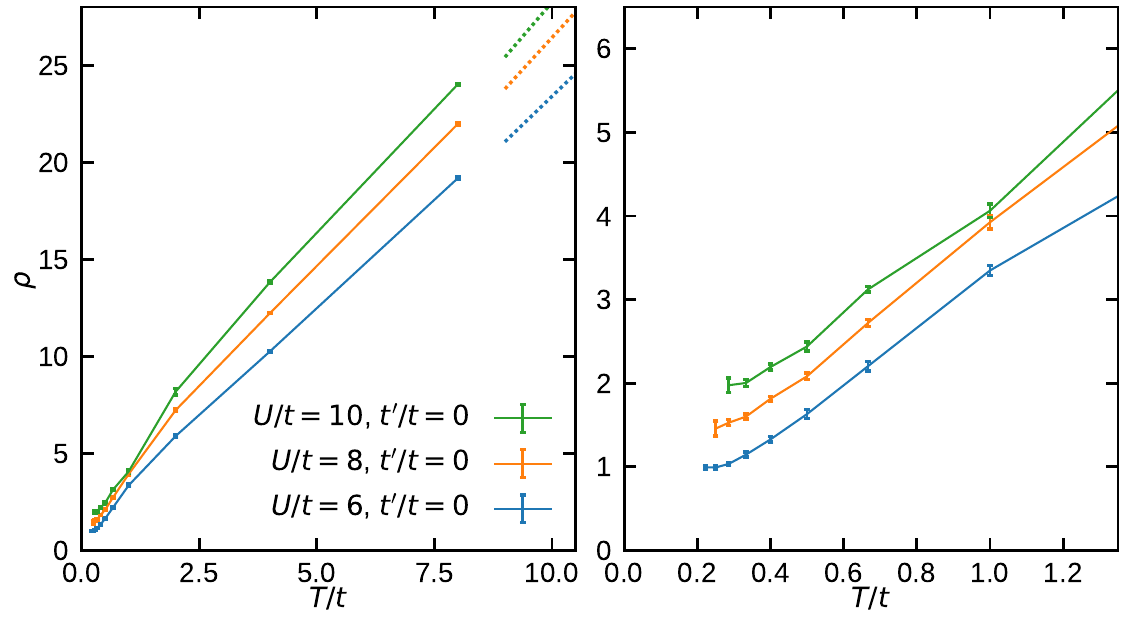}
\caption{DC resistivity extracted by maximum entropy analytic continuation for different interaction strengths $U/t$, all at hole doping $p=0.2$}
\label{fig:Udep}
\end{figure}

In Fig.~\ref{fig:Udep}, we plot our results for DC resistivity from analytic continuation for various choices of Hubbard model parameters. Consistent with expectations the resistivity rises with $U/t$, but the qualitative behavior is similar to that of the simulations presented in the main text.

\noindent\underline{Infinite temperature moments expansion}

In this section we set the nearest neighbor hopping $t = 1$ and use $t$ to denote real time.

The exact evaluation of equal time observables at infinite temperature is trivial due to the decoupling of densities. This is no longer true for unequal time correlators and response functions. However, the moments of response functions correspond to the coefficients of an Taylor series expansion of unequal time correlators at $t=0$ and hence can be evaluated exactly. Below we outline our approach for evaluating moments of the optical conductivity and constructing the current correlator and optical conductivity from its moments. Similar techniques have been applied to various problems in literature (see for instance \cite{Starykh1997,Lindner2010,Perepelitsky2016,Khait2016}).

Without loss of generality, we consider only the $x$ component of current and the $x x$ component of the conductivity tensor.
The optical conductivity is related to the current-current correlation function by
\begin{equation}
\sigma_1(\omega) = \frac{1-e^{-\beta\omega}}{2\omega} \int_{-\infty}^\infty dt e^{i \omega t} \Lambda(t).
\end{equation}
Below, we work only in the limit of infinite temperature.
\begin{align}
T \sigma_1(\omega) &= \frac{1}{2} \int_{-\infty}^\infty dt e^{i \omega t} \Lambda(t) \\
\Lambda(t) \equiv \llangle j(t) j \rrangle &= 2 T \int_{-\infty}^\infty \frac{d\omega}{2 \pi} e^{-i \omega t} \sigma_1(\omega). \label{jjs}
\end{align}
The $k$th moment of the optical conductivity is
\begin{equation}
\mu_k = \int_{-\infty}^\infty \frac{d\omega}{2 \pi} \omega^k \sigma(\omega).
\end{equation}
We focus on the real part $\sigma_1$ and hence consider only even moments. By (\ref{jjs}),
\begin{align}
\mu_{2 k} &= \frac{1}{2 T} \left(i \frac{d}{d t}\right)^{2k} \llangle j(t) j \rrangle \bigg\rvert_{t=0} \\
&= \frac{1}{2 T} \llangle (\mathcal{L}^{2 k} j) j \rrangle = \frac{1}{2 T} (-1)^{k} \llangle (\mathcal{L}^k j) (\mathcal{L}^k j) \rrangle. \label{ljlj}
\end{align}
Here, $\mathcal{L}$ is the Liouvillian, defined by $\mathcal{L} A = [H, A]$ for the operator $A$. In the last line, we have used $\llangle (\mathcal{L} A) B \rrangle = - \llangle A (\mathcal{L} B) \rrangle$.

The form of $\mathcal{L}^k j$ and its corresponding expectation value in (\ref{ljlj}) are determined algorithmically. We work in the thermodynamic limit and the evaluated moments are exact up to numerical error. For the 2d Hubbard model with parameters $U=6$, $t'=-0.25$, we calculate up to $k = 9$, obtaining all moments up to $\mu_{18}$. We also consider the parameters $U=6$, $t'=0$, for which we calculate up to $k = 11$ and $\mu_{22}$. Examples of moments are listed in Table \ref{tab_moments}.

\begin{table}
\centering
\begin{tabular}{|c|c|c|} 
\hline
k & $2 T \mu_k$, $t'=0$ & $2 T \mu_k$, $t'=-0.25$ \\
\hline
0 & 0.96 & 1.08 \\
2 & 16.5888 & 18.6624 \\
4 & 879.2064 & 972.23328 \\
6 & 71350.419456 & 79126.963776 \\
8 & 7957186.19136 & 8878803.683202207 \\
10 & 1161496143.3732295 & 1300988462.1862698 \\
12 & 214334017036.04272 & 240608065274.43448 \\
14 & 48564962187310.74 & 54707332165340.93 \\
16 & 13142108208577344 & 14920607763570232 \\
18 & 4134593753382283264 & 4764833870643252224 \\
20 & 1476890369651272056832 & \\
22 & 588420629083284729495552 & \\
\hline
\end{tabular}
\caption{Moments of the optical conductivity evaluated through (\ref{ljlj}), for the 2d Hubbard model with parameters $U = 6$, $p=0.2$ hole doping, and $t'$ indicated in the top row. Values are exact up to a relative accuracy $\sim 10^{-15}$ due to numerical precision. Kahan summation is used to minimize accumulated errors.}
\label{tab_moments}
\end{table}

Due to the rapid growth of the number of terms in $\mathcal{L}^k j$, the primary computational limitation is memory: we work on systems with 128GB of RAM, which can store expressions containing up to $\sim 10^9$ terms. For evaluating higher order moments, it is necessary to utilize hard drive storage and/or distribute the computation across multiple nodes. It would require enormous computational effort to significantly extend our current data, as the number of terms in $\mathcal{L}^k j$ increases by around an order of magnitude when incrementing $k$.

We consider two approaches to estimating the optical conductivity given the moments:

\begin{figure}
\centering
\includegraphics{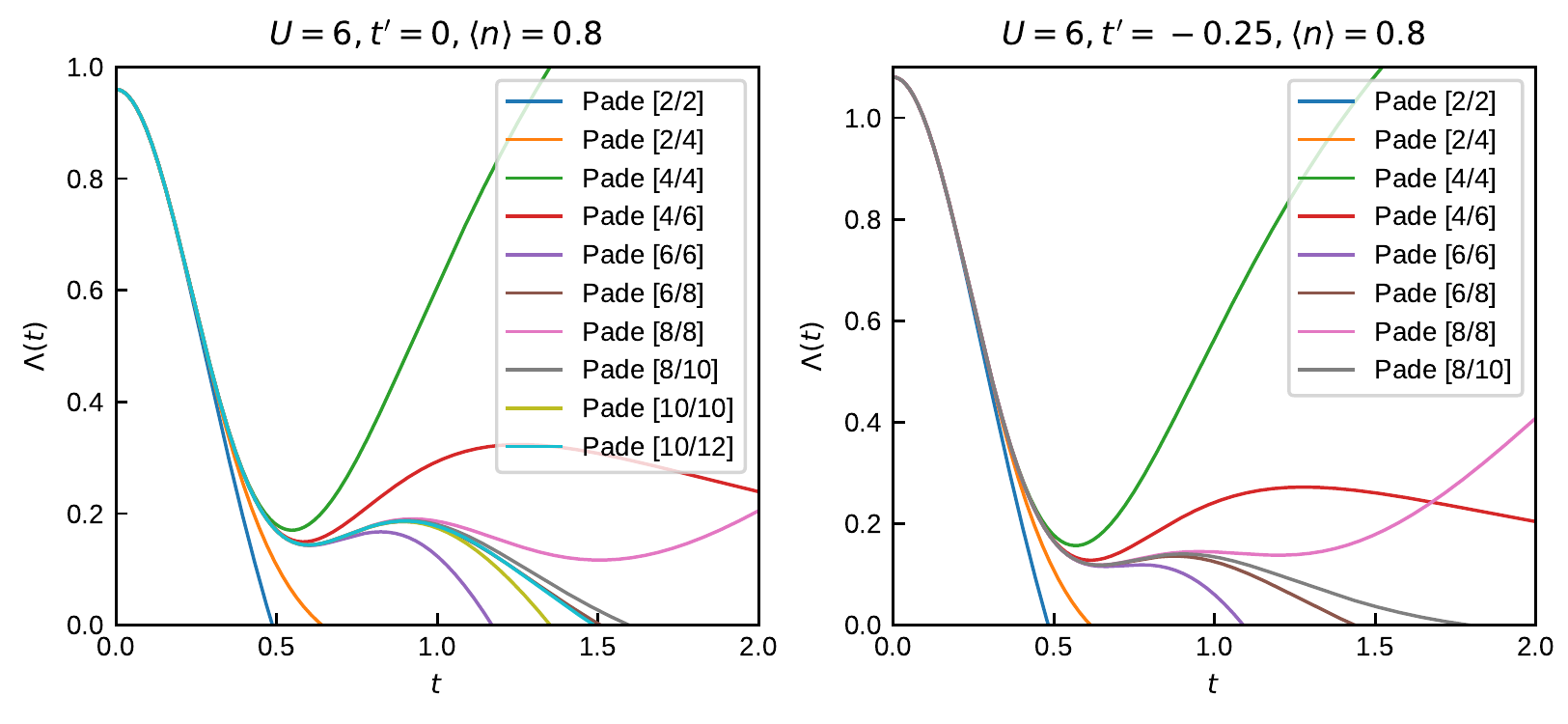}
\caption{Pad\'e approximants of various orders to $\Lambda(t)$.}
\label{fig:pade_lmbd}
\end{figure}

\begin{enumerate}
\item The moments provide the Taylor series coefficients of the current-current correlation function: $\Lambda(t) = \sum_{k=0}^\infty a_{2 k} t^{2 k}$, $a_{2 k} = 2 T (-1)^k \mu_{2 k}  / (2 k)!$. Note that (\ref{jjs}) implies $\Lambda(t)$ is even and real in the infinite temperature limit. These coefficients uniquely determine Pad\'e approximants of $\Lambda(t)$. Figure S\ref{fig:pade_lmbd} shows Pad\'e approximants of various orders. For our parameters, the highest order Pad\'e approximants are essentially converged up to time $t \approx 1$.

The behavior of Pad\'e approximants in the converged region $t \lesssim 1$ already provides considerable insight into the structure of the infinite temperature optical conductivity. First, the decay of current correlations is not monotonic, and reaches a local minimum near $t \approx 0.55$ and a local maximum near $t \approx 1$. This suggests oscillatory behavior with period $\approx 1$. Regardless of whether oscillations persist for $t \gtrsim 1$, the presence of a complete period of oscillation implies that there is a peak, possibly broad, in the optical conductivity around $\omega \approx$. Second, as the local minimum is still above $0$, there must be a significant contribution to the current correlation from a more slowly decaying function.

\begin{figure}
\centering
\includegraphics{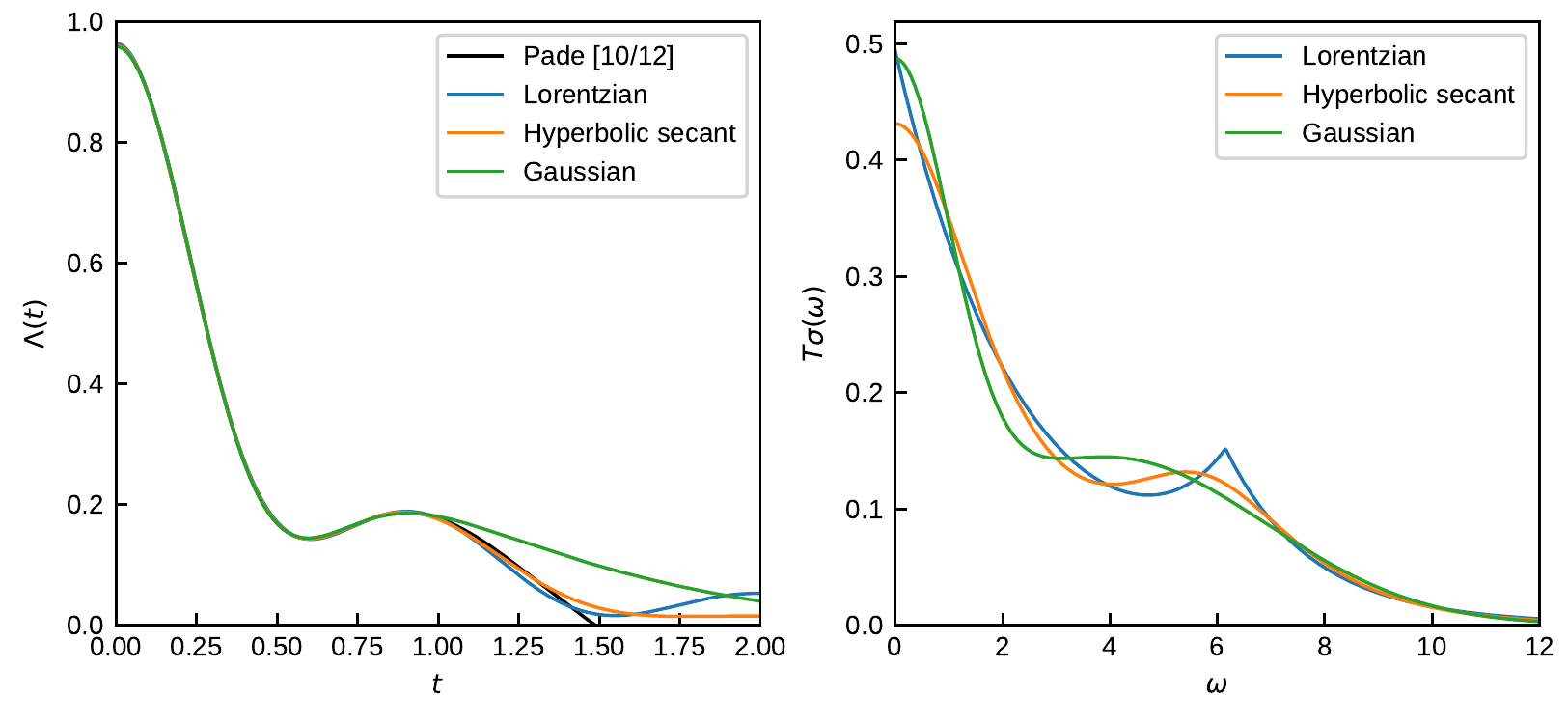}
\caption{Fits to the Pad\'e approximants of $\Lambda(t)$ for $U=6$, $t'=0$, and $\langle n \rangle = 0.8$. The functional form of the fit is $A_1 f(\Gamma_1 t) + A_2 f(\Gamma_2 t) \cos(\omega t)$, where $f$ is the function indicated in the figure legend and $A_1, A_2, \Gamma_1, \Gamma_2$, and $\omega$ are the parameters of the fit. A least-squares fit to the $[10/12]$ Pad\'e approximant for $0 < t < 1$ is performed.}
\label{fig:pade_fit}
\end{figure}

Based on the above considerations, a minimal form of the optical conductivity would be the sum of peaks at $\omega = 0$ and $\omega \approx \pm 6$. We consider peaks with Gaussian, hyperbolic secant, and exponentially decaying profiles, for a total of 9 possible profiles. These forms are Fourier transformed to time, and fitted against the highest order Pad\'e approximant of $\Lambda(t)$ for time $0 \leq t \leq 1$. The error of the fits are $\sim 10^{-3}$ in all cases. Plots of the fitted functions and corresponding optical conductivities are shown in Fig.~S\ref{fig:pade_fit}.

We briefly discuss our choice of peak profiles. First, all moments of the optical conductivity are finite. This is visible through (\ref{ljlj}): since both the Hamiltonian and the current operator are local, $\llangle (\mathcal{L}^{2 k} j) j \rrangle$ cannot diverge. Hence, at high frequency the optical conductivity must decay faster than any power law. This is the rationale for choosing profiles with exponential (or faster decaying) tails.

In \cite{Mukerjee2006,Kovtun2015}, for generic nonintegrable Hamiltonians, a nondivergent singularity at $\omega=0$ is predicted based on nonlinear coupling between energy and charge diffusive modes. The singularity has the form $\lim_{\omega \to 0} \sigma(\omega) = a - b \abs{\omega}^{d/2}$. In $d=1$ dimension, evidence for this was provided through exact diagonalization of a 1d model \cite{Mukerjee2006}. While similar behavior in 2d is plausible, no such direct numerical evidence currently exists, as exact diagonalization is limited to linear systems sizes of $\sim 4$, which would significantly round off any singularity \cite{Mukerjee2006}.

The exponentially decaying profile $e^{-\abs{\omega}}$ has a cusp consistent with the type of singularity expected in 2d. As both this sort of peak profile and nonsingular profiles produce reasonable fits to $\Lambda(t)$ in Fig.~S\ref{fig:pade_fit}, we cannot ascertain the existence of a singularity at $\omega = 0$. However, the fact that either type of peak profile results in similar looking optical conductivities suggests that even in the presence of a zero frequency cusp, the true value of $\sigma(\omega=0)$ is close to what's shown in Fig.~S\ref{fig:pade_fit}.

\item The optical conductivity may written as a continued fraction \cite{Starykh1997,Lindner2010,Perepelitsky2016,Khait2016}:
\begin{equation}
\sigma(\omega) = \cfrac{2 \mu_0}{i \omega + \cfrac{\abs{\Delta_1}^2}{i \omega + \cfrac{\abs{\Delta_2}^2}{i \omega + \dots}}} \label{cfinf}
\end{equation}
The moments $\mu_0, \mu_1, \dots, \mu_n$ exactly determine the recurrents $\abs{\Delta_1}^2, \dots, \abs{\Delta_n}^2$ through the following. Let $M_i^0 = M_i^1 = \mu_i / \mu_0$. $M_i^j$ for $j = 2, \dots, i$ is defined recursively by
\begin{equation}
M_i^j = \frac{M_i^{j-1}}{M_{j-1}^{j-1}} - \frac{M_{i-1}^{j-2}}{M_{j-2}^{j-2}}.
\end{equation}
The recurrents are given by the diagonal elements: $\abs{\Delta_i}^2 = M_i^i$.

\begin{figure}
\centering
\includegraphics{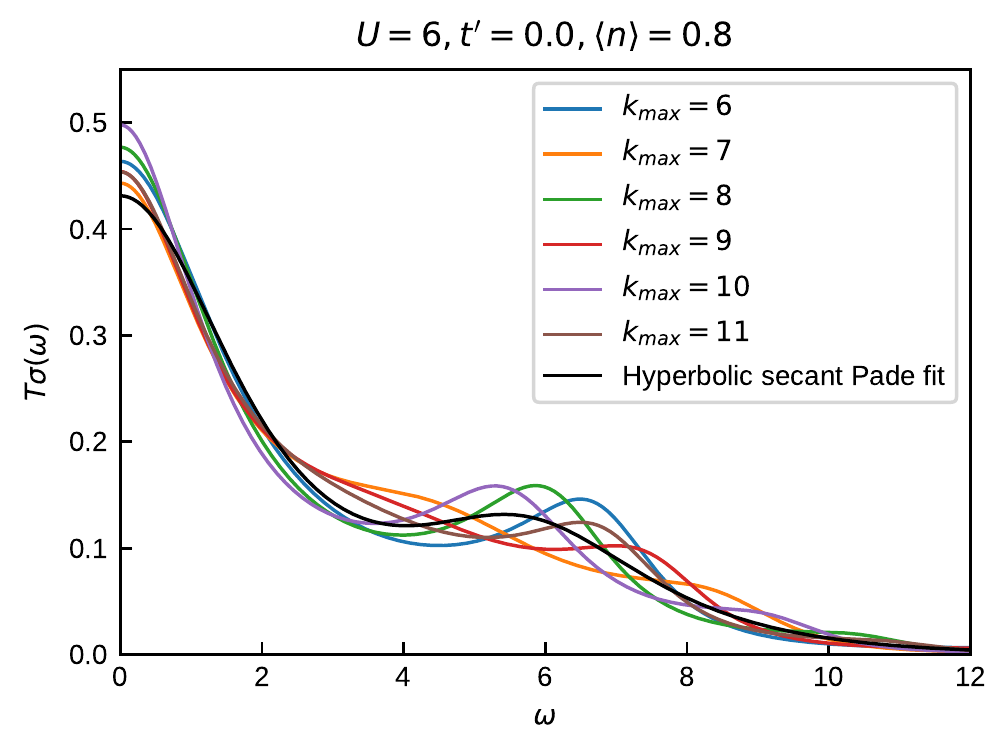}
\caption{Truncated continued fraction of $\Lambda(t)$ for $U=6$, $t'=0$, and $\langle n \rangle = 0.8$ as given in (\ref{cftrunc}). Also plotted is the hyperbolic secant fit to the highest order Pad\'e approximant.}
\label{fig:recc}
\end{figure}

Given the recurrents $\abs{\Delta_1}^2, \dots, \abs{\Delta_n}^2$, the task of constructing the optical conductivity is reduced to determining an appropriate truncation function $T(\omega)$:
\begin{equation}
\sigma(\omega) = \cfrac{2 \mu_0}{i \omega + \cfrac{\abs{\Delta_1}^2}{i \omega + \cfrac{\ddots}{i \omega + \cfrac{\abs{\Delta_n}^2}{i \omega + T(\omega)}}}} \label{cftrunc}
\end{equation}
This has been approached through various extrapolations in \cite{Starykh1997,Lindner2010,Khait2016}. A less sophisticated technique is given in \cite{Perepelitsky2016}, which amounts to setting the last fraction $\frac{\abs{\Delta_n}^2}{i \omega + T(\omega)} = \abs{\Delta_n}$. In Fig.~S\ref{fig:recc}, we show the results of applying this simple method.
\end{enumerate}

The qualitative and quantitative agreement between these completely different approaches provides confidence that the true form of the infinite temperature optical conductivity is unlikely to be considerably different from our estimates. We generally find that the first method of fitting to Pad\'e approximants, especially using hyperbolic secant profiles, to produce more robust results that tend to lie in the middle of the spread of spectra using truncated continued fractions. We thus use this fitted spectra in the data presented in the main text.

\noindent\underline{Average sign}

\begin{figure}
\centering
\includegraphics{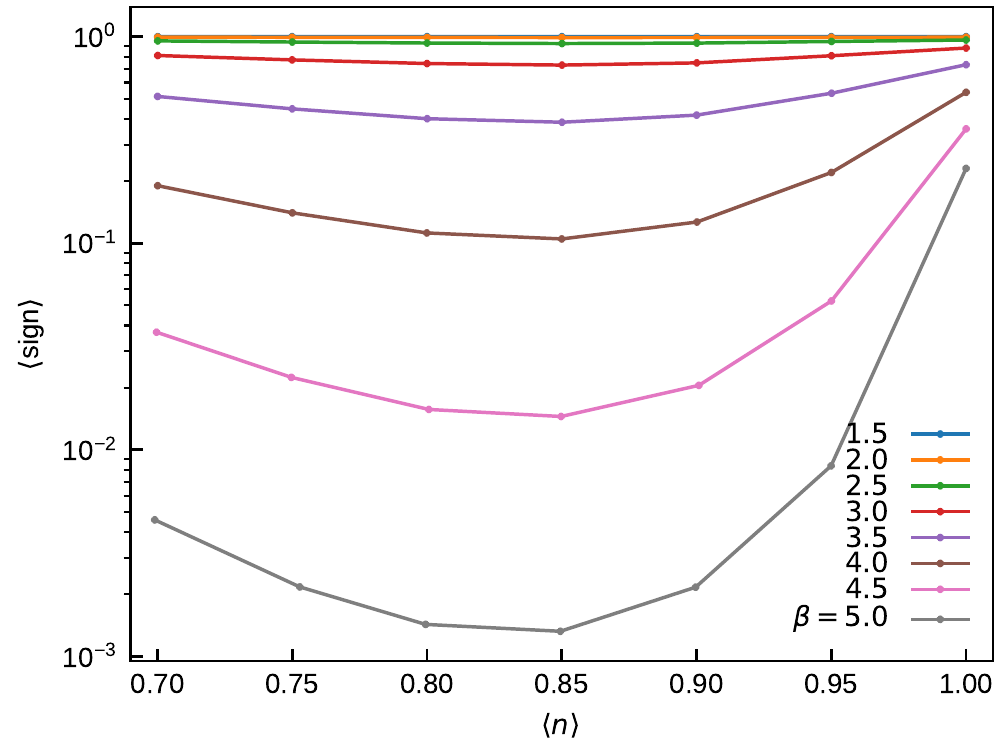}
\caption{Average sign of the DQMC simulations presented in the main text. Measured average sign is exactly $1$ for $\beta \leq 1.5$. Errors are smaller than symbol size: $\leq O(10^{-5})$ for the sign and $\leq O(10^{-4})$ for density.}
\label{fig:sign}
\end{figure}

DQMC simulations of the Hubbard model exhibit a sign problem upon doping and/or introducing a next nearest neighbor hopping. The average sign in our simulations is shown in Fig.~\ref{fig:sign}. As the sampling error is inversely proportional to the average sign, we consider and analyze carefully the random error in our results, especially at the lowest temperatures.

\noindent\underline{Sampling error analysis}

Here we describe the procedure used to estimate the random error of our data. In other words, we seek to describe the repeatability of our results: if we were to rerun our Monte Carlo simulations with different seeds, how different would be the final data each time? Because of the complexity of some of our data analysis, standard techniques of error propagation are inapplicable. We instead use the more general and powerful method of bootstrap resampling. For maximum fidelity, we resample at the beginning of data analysis, when the different bins of data are loaded, run our data analysis (e.g. the analytic continuation) on each set of resampled data, and then observe the bootstrap distribution of the final quantity of interest (e.g. optical conductivity). In the data presented in the main text, we use 1000 resamples and plot with errorbars representing $\pm 1$ standard error of the mean.

\begin{figure}
\centering
\includegraphics{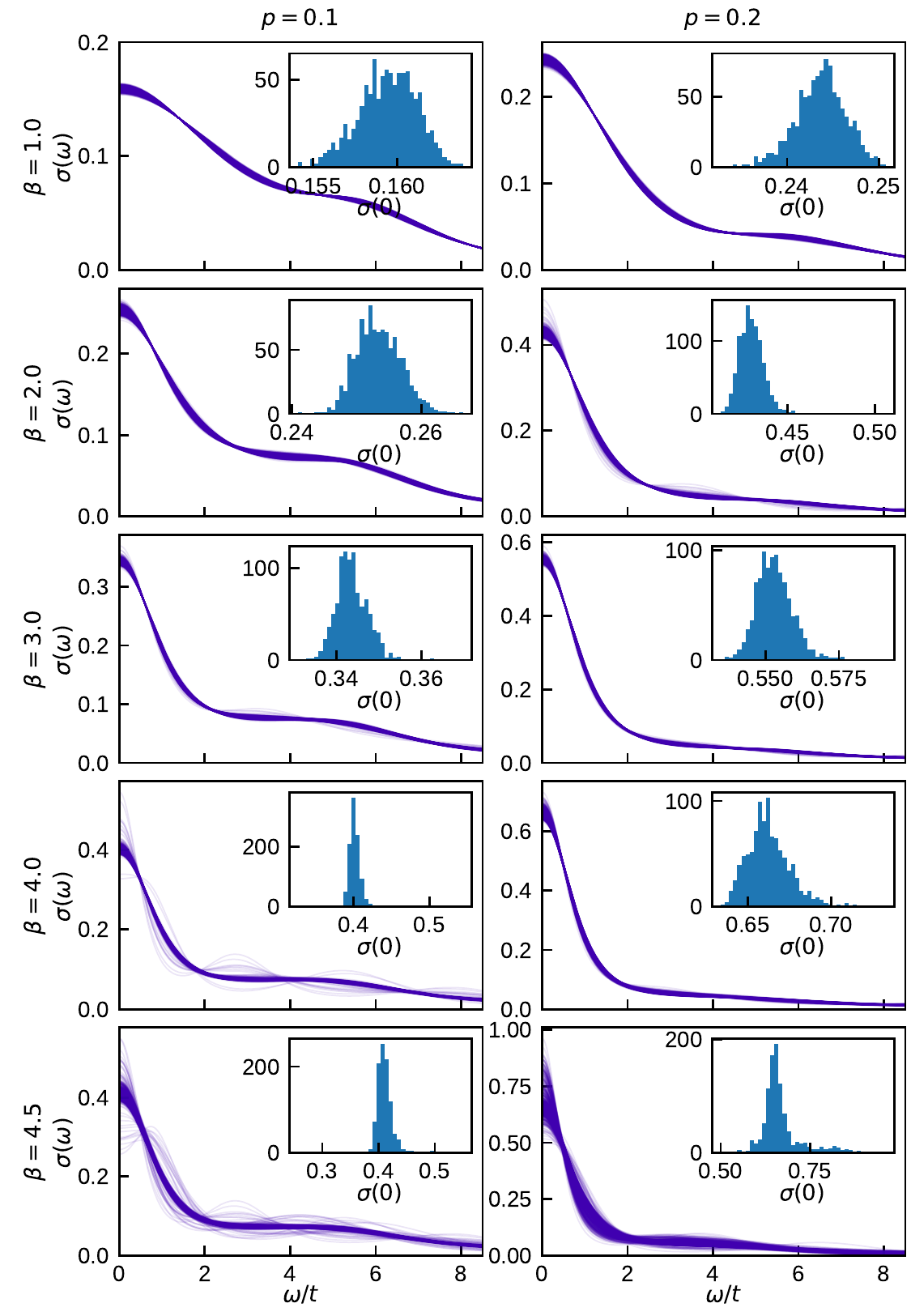}
\caption{Plots of optical conductivity obtained via analytic continuation of bootstrap resampled data. Each panel contains 1000 resamples. Insets: histograms of resampled DC conductivity.}
\label{fig:bs}
\end{figure}

In Fig.~\ref{fig:bs} we show examples of superposed plots of the optical conductivity where each curve corresponds to one bootstrap resample. Evidently, there is little variation between bootstrap resamples, except at low temperatures where sampling errors are large. The standard deviation of the bootstrap distribution of the DC resistivity is taken as the standard error of the mean shown by the errorbars of Fig.~\ref{fig:f2}.

\noindent\underline{Maximum entropy method: details and dependence on $\alpha$}

In analytic continuation of QMC data, we seek to invert the ill-conditioned linear equation
\begin{equation}
G(\tau) = \int d\omega K(\tau, \omega) A(\omega),
\end{equation}
where $G$ is an imaginary time Green's function measured in QMC, $K$ is the kernel, and $A$ is a spectral function. For optical conductivity, $\Lambda(\tau) = \int \frac{d\omega}{\pi} \frac{\omega e^{-\tau \omega}}{1-e^{-\beta \omega}} \sigma_1(\omega)$ and we take
\begin{align}
G(\tau) &= \frac{2 \Lambda(\tau)}{\Lambda(\omega=0)} \\
K(\tau, \omega) &= \frac{\omega (e^{-\tau \omega}+e^{-(\beta-\tau) \omega})}{1 - e^{-\beta \omega}} \\
A(\omega) &= \frac{2}{\pi\Lambda(\omega=0)} \frac{\Im \Lambda(\omega)}{\omega} = \frac{2}{\pi\Lambda(\omega=0)} \sigma_1(\omega),
\end{align}
where $\Lambda(\omega=0) = \int_0^\beta d\tau \Lambda(\tau)$. $A(\omega)$ is normalized to $1 = \int_0^\infty d\omega A(\omega)$.

The maximum entropy (MaxEnt) method of analytic continuation selects the optimal spectrum as the one which maximizes the functional $Q[A] = \alpha S - \chi^2/2$. $S[A] = -\int d\omega A(\omega) \log \frac{A(\omega)}{m(\omega)}$ is the entropy, such that $-S$ represents the amount of additional information in $A(\omega)$ relative to the model function $m(\omega)$. $\chi^2$ is a statistic quantifying the deviation of the reconstructed $G = K * A$ from the mean $G$ measured in QMC. Hence, MaxEnt selects a spectrum in agreement with the data, but with minimum additional information relative to the model function. The balance between these two factors is controlled by the parameter $\alpha$. The appropriate selection of $\alpha$ is important to the success of MaxEnt: large values may result in underfitting the data, giving a spectrum not consistent with the imaginary time data, whereas small values may lead to overfitting and spurious features. In general, and especially with large statistics, there is a range of appropriate $\alpha$ where these issues are avoided and the optimal $A(\omega)$ is quite insensitive to varying $\alpha$.

\begin{figure}
\centering
\includegraphics{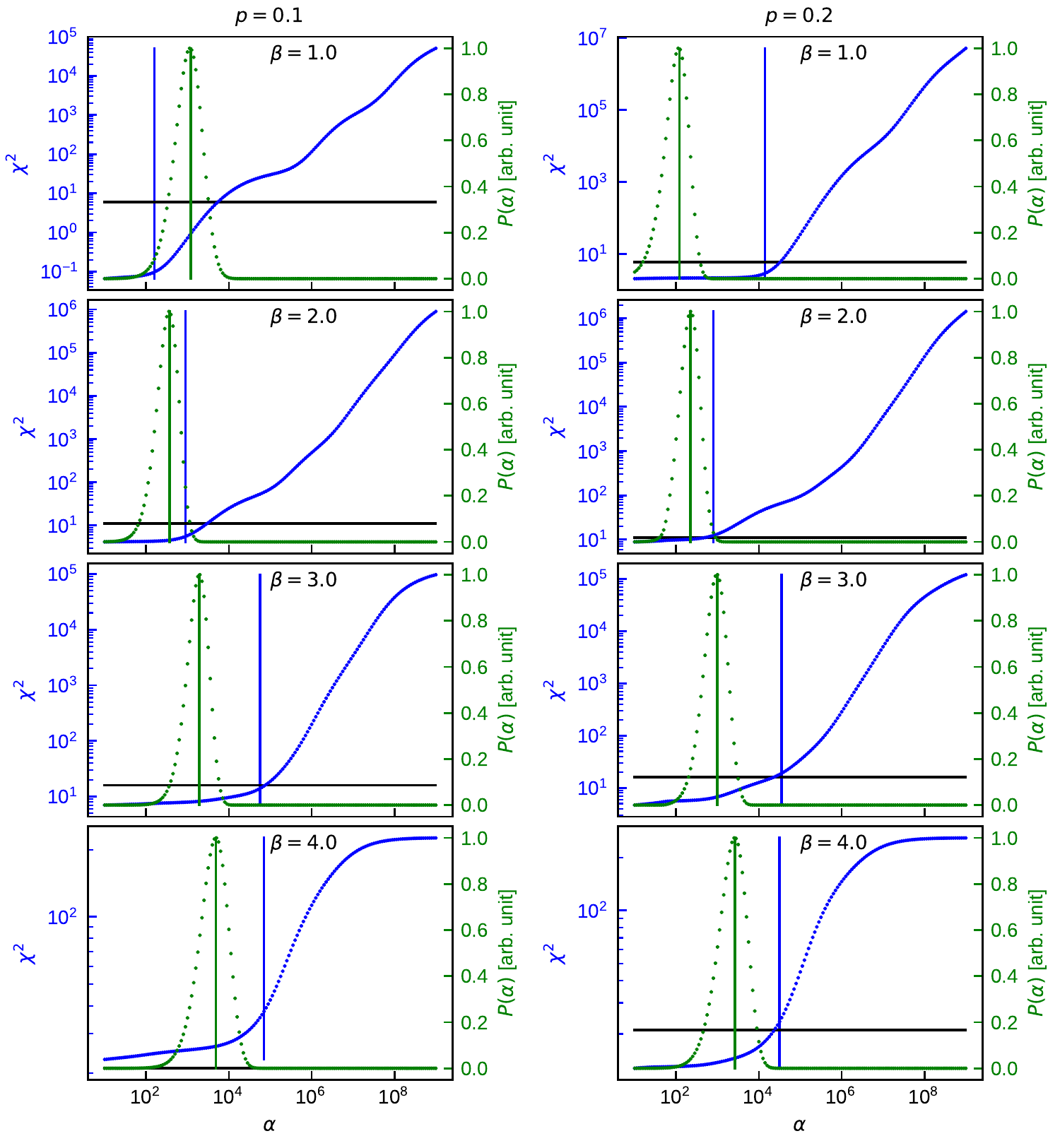}
\caption{Examples of diagnostic plots showcasing the selection of $\alpha$ in MaxEnt analytic continuation. Classic MaxEnt is defined by selecting the $\alpha$ maximizing the posterior probability $P(\alpha)$, as indicated by the green line. Historic MaxEnt is defined by taking the $\alpha$ for which $\chi^2$ equals the number of independent components in $G(\tau)$, indicated by the black horizontal line (i.e. the intersection of the black line with the blue dotted line gives $\alpha$). Bryan's method returns a weighted average of $A(\omega)$ with $P(\alpha)$ as the weight. A recently proposed method in Ref.~\cite{Bergeron2016} selects $\alpha$ as the location where a log-log plot of $\chi^2(\alpha)$ has maximal curvature; this selection is indicated by the blue line.}
\label{fig:diagn}
\end{figure}

\begin{figure}
\centering
\includegraphics{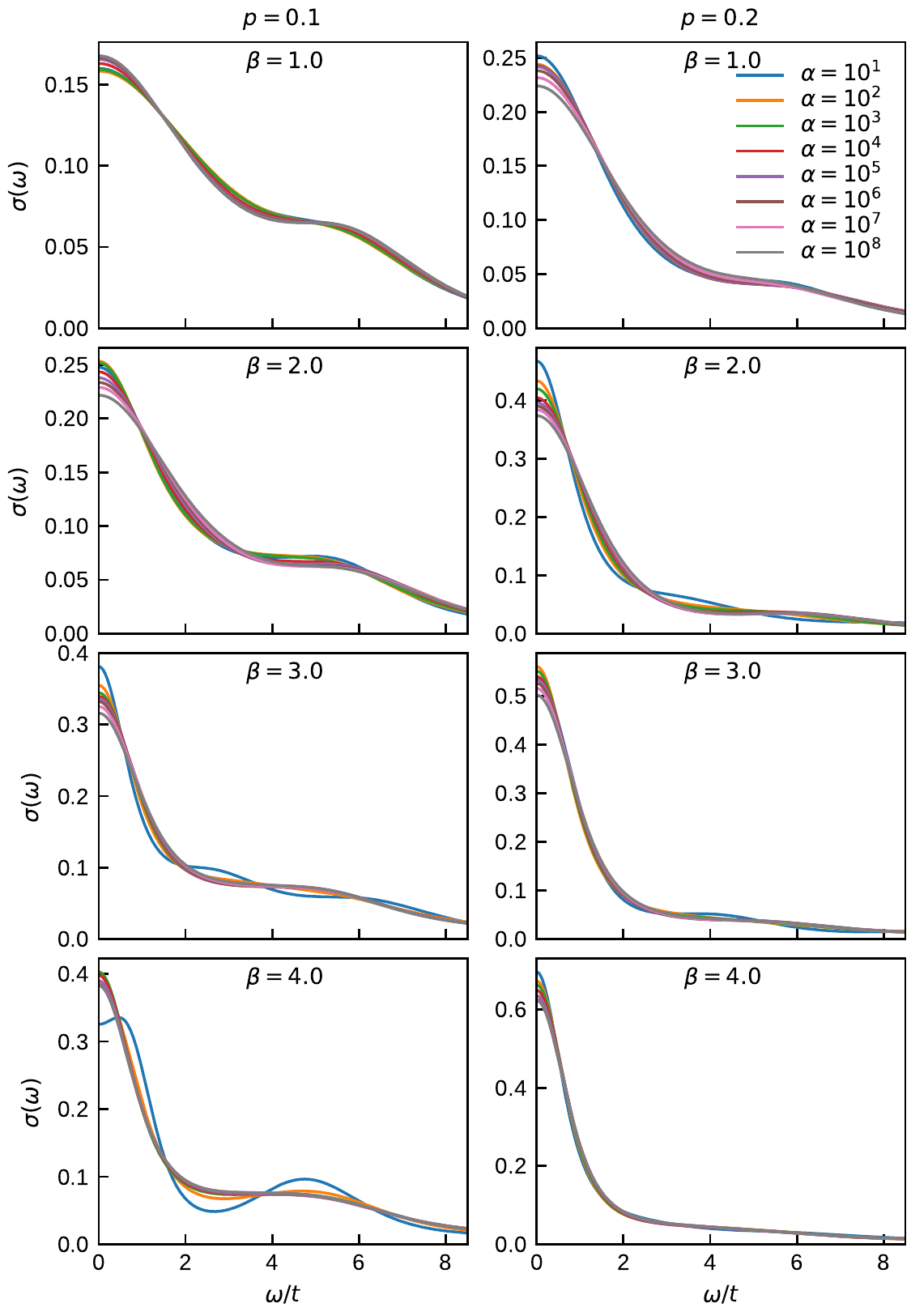}
\caption{Examples of optical conductivity from MaxEnt analytic continuation over a wide range of $\alpha$. In all cases, the spectra are almost insensitive to $\alpha$ near the optimal values of $\alpha$ selected by the various flavors of MaxEnt (see Fig.~\ref{fig:diagn}). Note that these plots encompass seven orders of magnitude of $\alpha$; typically the different flavors agree on the appropriate value of $\alpha$ to within one order of magnitude.}
\label{fig:alphas}
\end{figure}

Different procedures to select an appropriate $\alpha$ define the different flavors of MaxEnt. In Ref.~\cite{Jarrell1996}, three flavors are introduced: historic MaxEnt, classic Maxent, and Bryan's method. In historic MaxEnt, $\alpha$ is selected such that the resultant $\chi^2$ equals the independent degrees of freedom in the imaginary time QMC data. Alternatively, using Bayesian methods, it is possible to define a posterior probability $P(\alpha)$. Classic MaxEnt selects $\alpha$ as the one which maximizes $P(\alpha)$. Bryan's method \footnote{Not to be confused with Bryan's algorithm, which is an algorithm for finding the optimal $A(\omega)$ for a given $\alpha$ \cite{Jarrell1996}. In our MaxEnt code, regardless of the flavor of MaxEnt used, we always employ Bryan's algorithm.} performs a weighted average such that the final spectrum is $\int d\alpha P(\alpha) A_\alpha(\omega)$, where $A_\alpha(\omega)$ is the optimal spectrum for a given $\alpha$. One caveat to the latter two methods is that the estimation of $P(\alpha)$ is inaccurate when the model function is significantly different from the final spectrum; hence, Ref.~\cite{Bergeron2016} proposes to select $\alpha$ as the location of maximum curvation in a log-log plot of $\chi^2(\alpha)$.

In Fig.~\ref{fig:diagn} we plot examples of diagnostic data to show the selection of $\alpha$ using these four different flavors. Despite variations in the choice of $\alpha$ between these methods, we show in Fig.~\ref{fig:alphas} that the final resultant $\sigma(\omega)$ is insensitive to the value of $\alpha$ over a significant range. Thus, while ultimately we choose the classic formulation of MaxEnt, our data are of sufficient quality that all variants of MaxEnt give nearly identical results.

\noindent\underline{Systematic error of analytic continuation}

In the previous sections we have shown that our analytic continuation procedure using MaxEnt is precise and repeatable, with sampling errors small enough to be qualitatively unimportant to our results. Uncertainty remains regarding systematic biases resulting from analytic continuation. While performing analytic continuation exactly is in general impossible in the presence of noise, we can assess the accuracy of our results and gain insights into the capabilities and limitations of MaxEnt via the following: considering various test spectra, transforming to imaginary time, performing analytic continuation back to real frequency, and comparing against the original spectra. After transforming the test spectra to imaginary time, we add noise sampled from the same multivariate Gaussian distribution as that in the Monte Carlo calculations. Furthermore in the analytic continuation, we also use the full covariance matrix of $G(\tau)$ estimated by the Monte Carlo.

\begin{figure}
\centering
\includegraphics{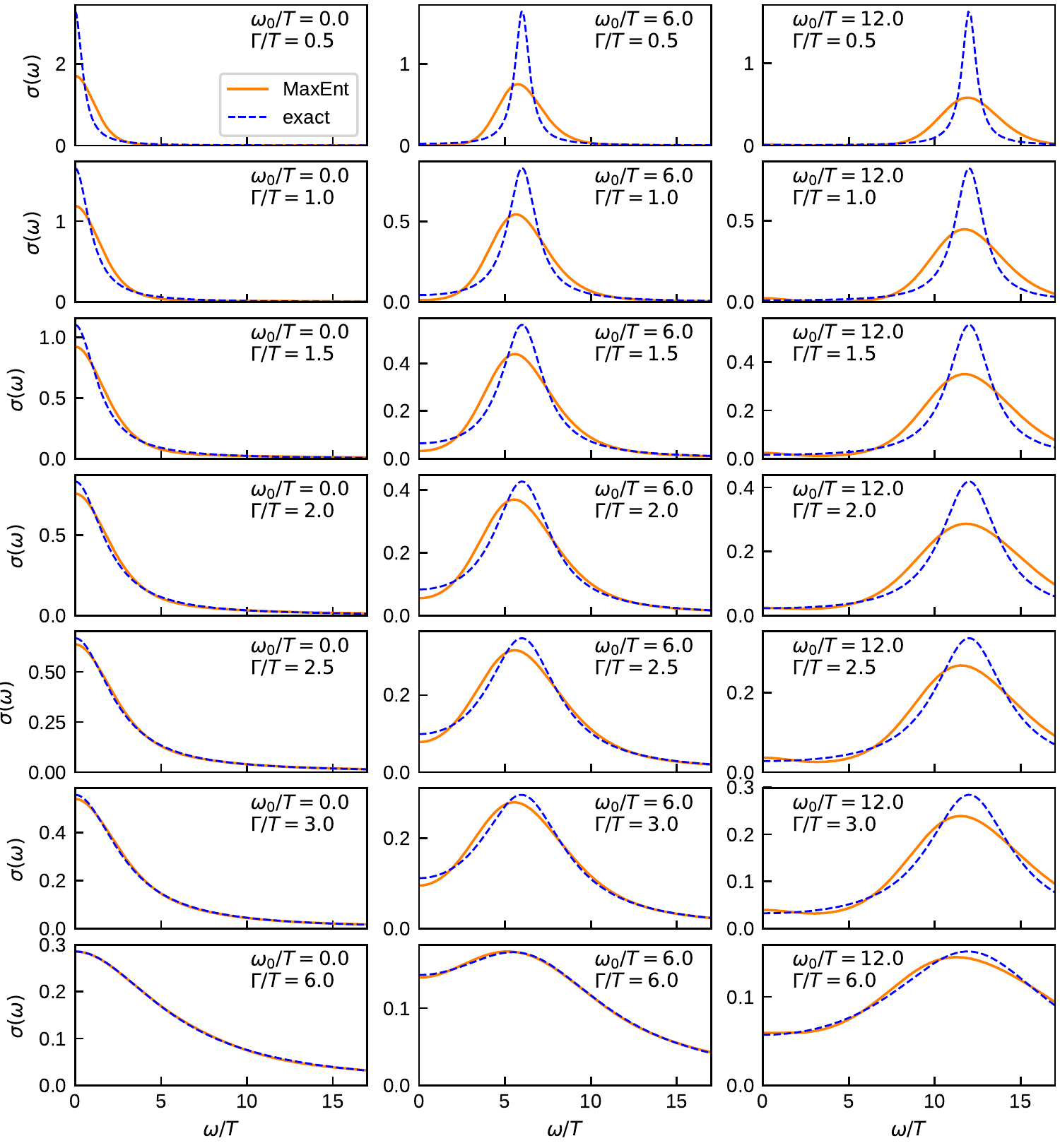}
\caption{Tests of MaxEnt analytic continuation. The exact spectrum is Lorentzian, symmetrized to be an even function: $\sigma(\omega) \propto \frac{1}{(\omega-\omega_0)^2 + \Gamma^2} + \frac{1}{(\omega+\omega_0)^2 + \Gamma^2}$. The analytic continuation uses the same covariance matrix as from the DQMC simulations at $p=0.2$, $\beta=3.0$. (This choice is unimportant except for $\beta \gtrsim 4.5$ for which the sampling errors can be large.) The model function is the same as in the MaxEnt calculation in the main text: related to $\sigma(\omega)$ at the next highest temperature $\beta=2.5$. Using a flat model function yields essentially identical results. Since these model functions are significantly different from the test spectra, the estimation of the posterior probability $P(\alpha)$ is inaccurate and so we use the method in Ref.~\cite{Bergeron2016} to select $\alpha$ instead of classic MaxEnt.}
\label{fig:recon}
\end{figure}

\begin{figure}
\centering
\includegraphics{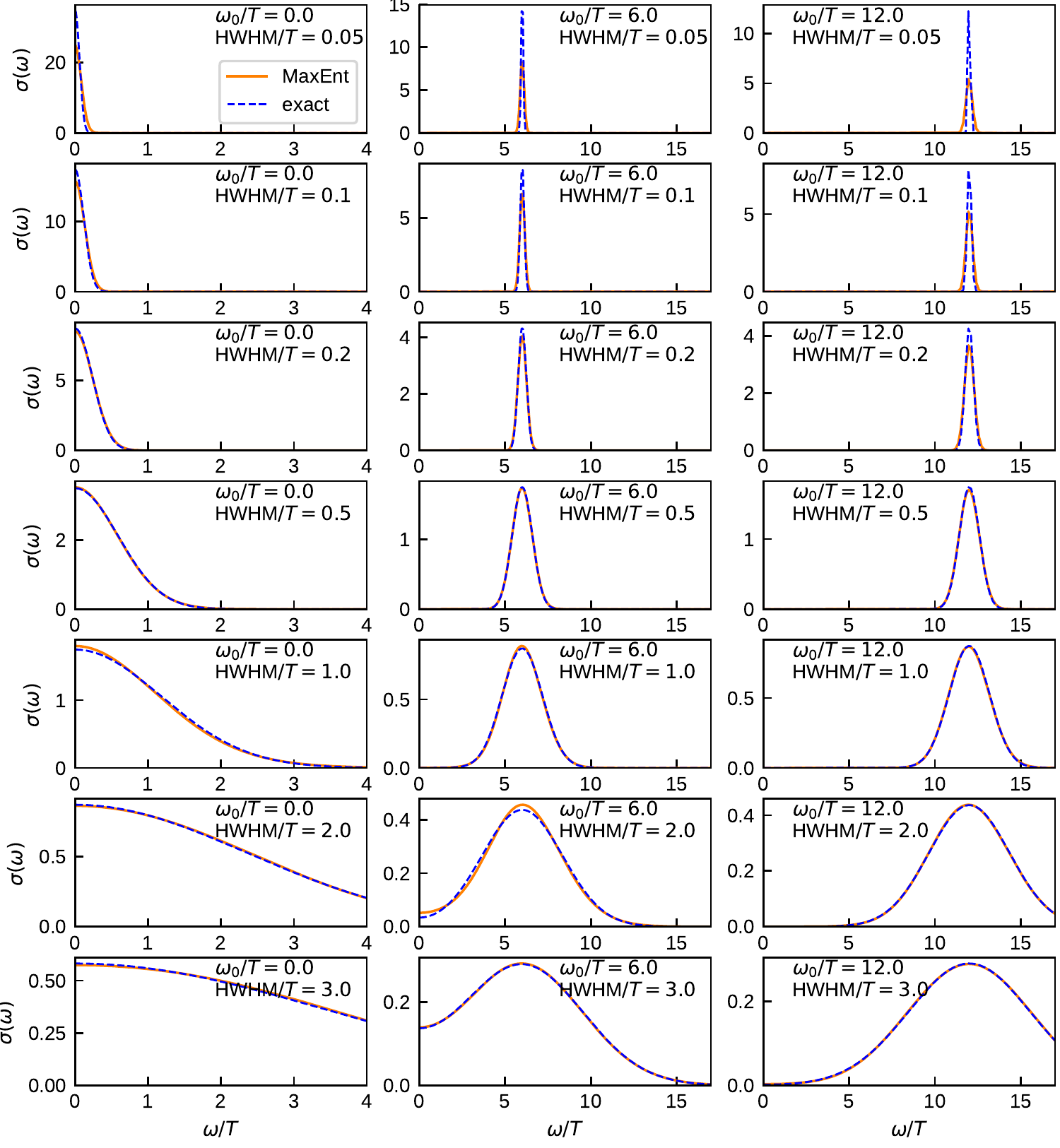}
\caption{Same as Fig.~\ref{fig:recon}, expect with Gaussian test spectra: $\sigma(\omega) \propto e^{-(\omega-\omega_0)^2/(2\sigma^2)} + e^{-(\omega+\omega_0)^2/(2\sigma^2)}$ with $\mathrm{HWHM} = \sqrt{2 \log 2} \sigma$.}
\label{fig:recon_gauss}
\end{figure}

With this method we evaluate the intrinsic limitations of analytic continuation of imaginary time data at finite temperature. A common misconception is that analytic continuation is inaccurate for frequencies $\omega \lesssim T$ but accurate for $\omega \gtrsim T$. Actually, the limitations of analytic continuation is that of resolution: peaks narrower than $\sim T$ or multiple peaks with separation within $\sim T$ tend to be blurred. In fact, this effect becomes more pronounced at higher frequencies, as demonstrated in Fig.~\ref{fig:recon}. Heuristically, this occurs because small real frequencies are closer to the imaginary axis Matsubara frequencies.

To quantify the amount of blurring that occurs, we try spectra containing peaks of various widths in Fig.~\ref{fig:recon}. When the half width at half maximum $\gtrsim 2T$, the analytically continued spectra is very similar or nearly identical to the original test spectra. Therefore the effect of analytic continuation cannot be reduced to an indiscriminate blurring; only features narrower than $\sim T$ are affected and broader features can be reproduced very accurately. Furthermore, we see from this that significant blurring can be diagnosed by the presence of peaks with HWHM $\lesssim 2T$ in the analytically continued spectra. In our analytically continued DQMC data, the zero frequency peaks tend to have HWHM $\sim 3 T$ even for the most narrow peaks (e.g. at the highest considered hole dopings).

In this analysis, we primarily focus on exact spectra contain a Lorentzian peak. In Fig.~\ref{fig:recon_gauss}, we also consider Gaussian peaks. Here, the performance of MaxEnt is remarkable: even peaks with width an order of magnitude below temperature can be reproduced fairly accurately. The main difference between Gaussian and Lorentzian profiles is the heavier tails in the latter. We found in Fig.~\ref{fig:f1} that in general the tails of the Drude peak are significant and blend into the Hubbard peak at $\omega \sim U$; hence our analysis with Lorentzian profiles is more relevant. We have also considered including a Hubbard peak in the test spectra, and find little change in our results.

\begin{figure}
\centering
\includegraphics{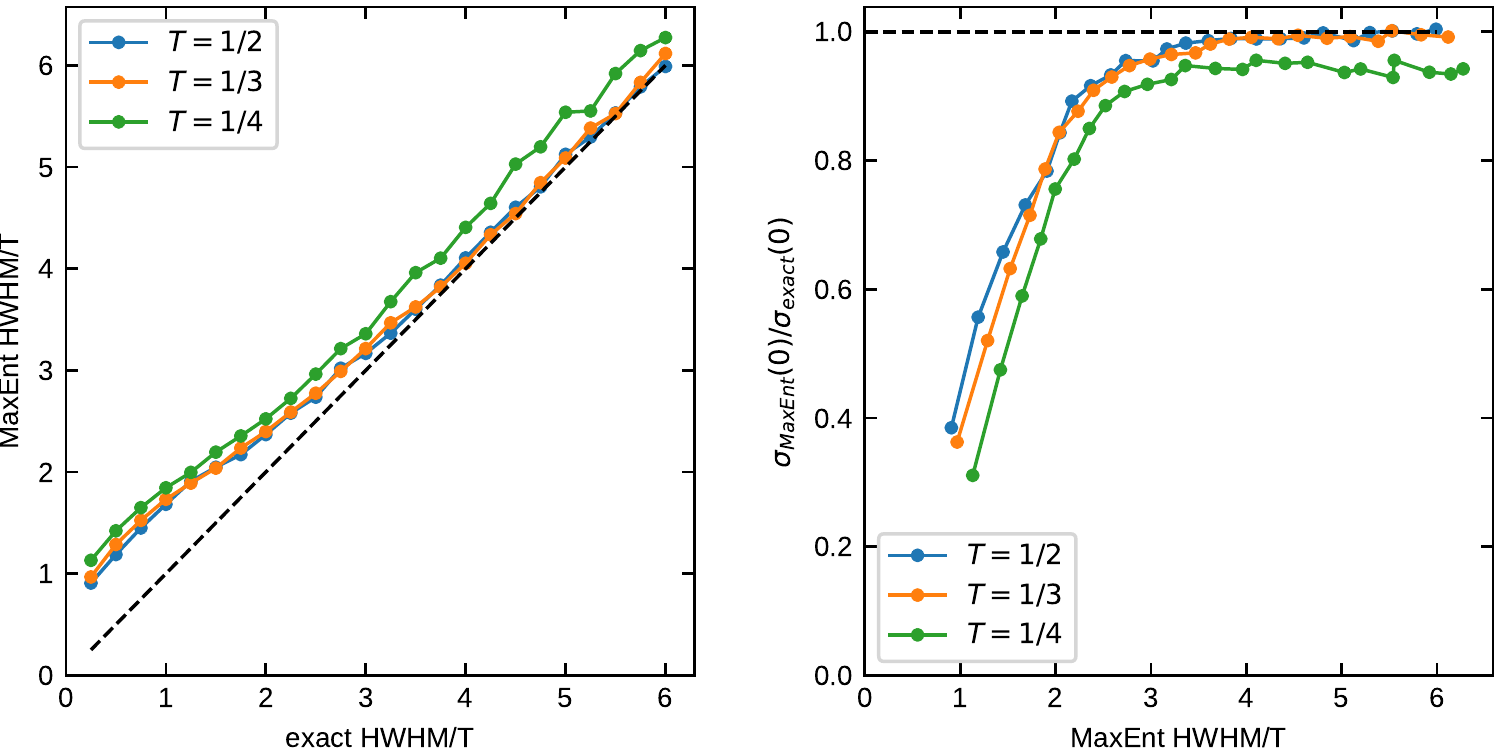}
\caption{Analysis of results similar to that in Fig.~\ref{fig:recon}. The different lines are for covariance matrices from DQMC simulations at different temperatures with doping $p=0.2$. Dashed lines represent the ideal results.}
\label{fig:hwhm}
\end{figure}

In Fig.~\ref{fig:hwhm}, we focus on the case of a Lorentzian peak at $\omega=0$ and plot the HWHM of the analytically continued spectra against that of the test spectra. As seen already in Fig.~\ref{fig:recon}, broadening and blurring of the peak decreases when the true HWHM becomes $\gtrsim 2.5$. We also plot in Fig.~\ref{fig:hwhm} the ratio of DC conductivities between the analytically continued and test spectra against the HWHM of the analytically continued spectra. Here we see that when the peak HWHM of the analytically continued spectra exceeds around $2T$ or $2.5T$, the inaccuracy of the DC conductivity is $\sim 10\%$. In our analytically continued DQMC data, the zero frequency peaks tend to have HWHM $\sim 3 T$ even for the most narrow peaks (e.g. at the highest considered hole dopings), placing our results in the regime where analytically continued data is trustworthy up to $\sim 10\%$. Based on all of the above analysis, we believe that our analytically continued DQMC data for DC resistivity is affected by systematic inaccuracies of analytic continuation by a few to several percent. This is comparable to the sampling error and sufficiently small that our conclusions are not affected.


\begin{thebibliography}{10}

\bibitem{Fradkin2015}
E.~Fradkin, S.~A. Kivelson, J.~M. Tranquada, {\it Rev. Mod. Phys.\/} {\bf 87},
  457 (2015).

\bibitem{Keimer2015}
B.~Keimer, S.~A. Kivelson, M.~R. Norman, S.~Uchida, J.~Zaanen, {\it Nature\/}
  {\bf 518}, 179 (2015).

\bibitem{Emery1995}
V.~J. Emery, S.~A. Kivelson, {\it Phys. Rev. Lett.\/} {\bf 74}, 3253 (1995).

\bibitem{Gunnarsson2003}
O.~Gunnarsson, M.~Calandra, J.~E. Han, {\it Rev. Mod. Phys.\/} {\bf 75}, 1085
  (2003).

\bibitem{Hussey2004}
N.~E. Hussey, K.~Takenaka, H.~Takagi, {\it Philosophical Magazine\/} {\bf 84},
  2847 (2004).

\bibitem{BSS}
R.~Blankenbecler, D.~J. Scalapino, R.~L. Sugar, {\it Phys. Rev. D\/} {\bf 24},
  2278 (1981).

\bibitem{White1989}
S.~R. White, {\it et~al.\/}, {\it Phys. Rev. B\/} {\bf 40}, 506 (1989).

\bibitem{Khait2016}
I.~Khait, S.~Gazit, N.~Y. Yao, A.~Auerbach, {\it Phys. Rev. B\/} {\bf 93},
  224205 (2016).

\bibitem{Starykh1997}
O.~A. Starykh, A.~W. Sandvik, R.~R.~P. Singh, {\it Phys. Rev. B\/} {\bf 55},
  14953 (1997).

\bibitem{Lindner2010}
N.~H. Lindner, A.~Auerbach, {\it Phys. Rev. B\/} {\bf 81}, 054512 (2010).

\bibitem{Perepelitsky2016}
E.~Perepelitsky, {\it et~al.\/}, {\it Phys. Rev. B\/} {\bf 94}, 235115 (2016).

\bibitem{Jarrell1996}
M.~Jarrell, J.~E. Gubernatis, {\it Phys. Rep.\/} {\bf 269}, 133 (1996).

\bibitem{Gunnarsson2010}
O.~Gunnarsson, M.~W. Haverkort, G.~Sangiovanni, {\it Phys. Rev. B\/} {\bf 82},
  165125 (2010).

\bibitem{supp}
See supplementary materials.

\bibitem{Mukerjee2006}
S.~Mukerjee, V.~Oganesyan, D.~Huse, {\it Phys. Rev. B\/} {\bf 73}, 035113
  (2006).

\bibitem{Trivedi1996}
N.~Trivedi, R.~T. Scalettar, M.~Randeria, {\it Phys. Rev. B\/} {\bf 54}, R3756
  (1996).

\bibitem{Lederer2017}
S.~Lederer, Y.~Schattner, E.~Berg, S.~A. Kivelson, {\it Proc. Natl. Acad. Sci.
  (U.S.A.)\/} {\bf 114}, 4905 (2017).

\bibitem{Hartnoll2015}
S.~A. Hartnoll, {\it Nat. Phys.\/} {\bf 11}, 54 (2015).

\bibitem{Hartman2017}
T.~Hartman, S.~A. Hartnoll, R.~Mahajan, {\it Phys. Rev. Lett.\/} {\bf 119},
  141601 (2017).

\bibitem{Kokalj2017}
J.~Kokalj, {\it Phys. Rev. B\/} {\bf 95}, 041110 (2017).

\bibitem{Mousatov2018}
C.~H. Mousatov, I.~Esterlis, S.~Hartnoll, {\it arXiv:1803.08054\/}  (2018).

\bibitem{Prushke1995}
T.~Pruschke, M.~Jarrell, J.~Freericks, {\it Advances in Physics\/} {\bf 44},
  187 (1995).

\bibitem{Bergeron2011}
D.~Bergeron, V.~Hankevych, B.~Kyung, A.-M.~S. Tremblay, {\it Phys. Rev. B\/}
  {\bf 84}, 085128 (2011).

\bibitem{Deng2013}
X.~Deng, {\it et~al.\/}, {\it Phys. Rev. Lett.\/} {\bf 110}, 086401 (2013).

\bibitem{Xu2013}
W.~Xu, K.~Haule, G.~Kotliar, {\it Phys. Rev. Lett.\/} {\bf 111}, 036401 (2013).

\bibitem{Xu2016}
W.~Xu, W.~McGehee, W.~Morong, B.~DeMarco, {\it arXiv:1606.06669\/}  (2016).

\bibitem{Brown2018}
P.~T. Brown, {\it et~al.\/}, {\it arXiv:1802.09456\/}  (2018).

\bibitem{Nichols2018}
M.~A. Nichols, {\it et~al.\/}, {\it arXiv:1802.10018\/}  (2018).

\bibitem{Jiang2018}
H.-C. Jiang, T.~P. Devereaux, {\it arXiv:1806.01465\/}  (2018).

\bibitem{Zheng2017}
B.-X. Zheng, {\it et~al.\/}, {\it Science\/} {\bf 358}, 1155 (2017).

\bibitem{Huang2018}
E.~W. Huang, C.~B. Mendl, H.-C. Jiang, B.~Moritz, T.~P. Devereaux, {\it npj
  Quant. Mat.\/} {\bf 3}, 22 (2018).

\bibitem{Tomas2012}
A.~Tomas, C.~C. Chang, R.~Scalettar, Z.~Bai, {\it 2012 IEEE 26th International
  Parallel and Distributed Processing Symposium\/} (2012), pp. 308--319.

\bibitem{Jiang2016}
C.~Jiang, Z.~Bai, R.~Scalettar, {\it 2016 IEEE International Parallel and
  Distributed Processing Symposium (IPDPS)\/} (2016), pp. 463--472.

\bibitem{Kovtun2015}
P.~Kovtun, {\it Journal of Physics A: Mathematical and Theoretical\/} {\bf 48},
  265002 (2015).

\bibitem{Bergeron2016}
D.~Bergeron, A.-M.~S. Tremblay, {\it Phys. Rev. E\/} {\bf 94}, 023303 (2016).

\end{thebibliography}
\end{document}